# The carbon monoxide-rich interstellar comet 2I/Borisov

D. Bodewits[1], J. W. Noonan[2], P. D. Feldman[3], M. T. Bannister[4], D. Farnocchia[5], W. M. Harris[2], J.-Y. Li (李荐扬)[6], K. E. Mandt[7], J. Wm. Parker[8], Z. Xing (邢泽曦)[1,9]

[1]Physics Department, Auburn University, Leach Science Center, Auburn, AL 36849, USA; [2]Lunar and Planetary Laboratory, University of Arizona, 1629 E University Boulevard, Tucson, AZ 85721, USA; [3]Department of Physics and Astronomy, The Johns Hopkins University, 3400 N. Charles Street, Baltimore, MD 21218, USA; [4]School of Physical and Chemical Sciences — Te Kura Matū, University of Canterbury, Private Bag 4800, Christchurch 8140, New Zealand; [5]Jet Propulsion Laboratory, California Institute of Technology, 4800 Oak Grove Dr., Pasadena, CA 91109, USA; [6]Planetary Science Institute, 1700 E. Ft. Lowell Road, Suite 106, Tucson, AZ 85719, USA; [7]Johns Hopkins Applied Physics Laboratory, 11100 Johns Hopkins Rd., Laurel, MD 20723, USA; [8]Department of Space Studies, Southwest Research Institute, Suite 300, 1050 Walnut Street, Boulder, CO 80302, USA; [9]Department of Physics and Laboratory for Space Research, The University of Hong Kong, Pokfulam Road, Hong Kong SAR, China.

**Interstellar comets offer direct samples of volatiles from distant protoplanetary disks. 2I/Borisov is the first notably active interstellar comet discovered in our solar system[1]. Comets are condensed samples of the gas, ice, and dust that were in a star's protoplanetary disk during the formation of its planets and inform our understanding on how chemical compositions and abundances vary with distance from the central star. Their orbital migration moves volatiles[2], organic material, and prebiotic chemicals in their host system[3]. In our solar system, hundreds of comets have been observed remotely, and a few have been studied up close by space missions[4]. However, knowledge of extrasolar comets has been limited to what could be gleaned from distant, unresolved observations of cometary regions around other stars, with only one detection of carbon monoxide[5]. Here we report that the coma of 2I/Borisov contains significantly more CO than $H_2O$ gas, with abundances of at least 173%, more than three times higher than previously measured for any comet in the inner (<2.5 au) solar system[4]. Our ultraviolet observations of 2I/Borisov provide the first glimpse into the ice content and chemical composition of the protoplanetary disk of another star that is substantially different from our own.**

2I/Borisov is the second interstellar object discovered in our solar system, and it showed comet-like activity when it was discovered, indicating sublimating ices. Its brightness and location in the sky made it observable for months. This situation is very different from that of the first discovered interstellar object, 1I/'Oumuamua, which was much fainter, was visible for only a few weeks for most observers, and showed no detectable levels of gas or dust[6]. 2I/Borisov's outgassing makes it possible to probe the chemical composition of volatiles stored within the nucleus. We report on its observation by the Cosmic Origins Spectrograph (COS) on the Hubble Space Telescope (HST) during four epochs between December 11, 2019, and January 13, 2020 (Extended Data Figure 1).

An example of the HST/COS spectra is shown in Figure 1. In each of the four epochs, we clearly detected the emissions of several bands of the CO Fourth Positive system, excited by solar fluorescence[7] (Extended Data Figure 2). The derived CO production rates range between 6.4 to 10.7 x $10^{26}$ molecules/s, equivalent to 30 – 50 kg/s. The volatile content of cometary nuclei is generally considered to be a mixture of mostly $H_2O$, $CO_2$, CO, and a small percentage of other molecules[4]. Chemical abundances in the coma are often expressed relative to $H_2O$. In Figure 2, we compare our measurements of the CO production rates with $H_2O$ production rates[8,9,10]. Reported water production rates peak around perihelion on Dec. 8, 2019, when 2I/Borisov came within 2.006 au of the Sun and decreased rapidly afterwards[9]. In contrast, we observed that the CO production rates remained constant during our observations, with a possible maximum around December 30, 2019. Interpolating between measurements acquired with the Ultraviolet-Optical Telescope on board the Neil Gehrels Swift observatory (Swift/UVOT) indicate that even just one week after perihelion, the production rate of CO exceeded that of $H_2O$. Swift/UVOT and HST/COS observed 2I/Borisov simultaneously on Dec. 19 – 22, 2019 and on Jan. 13, 2020, and found CO to $H_2O$ abundances of 130 – 155%. This is a much higher CO/$H_2O$ abundance than has been measured in any comet in the inner solar system (<2.5 au), which mostly ranges between 0.2 – 23% with typical values around 4%[4].

CO ice is highly volatile and sublimates around 20 K in protoplanetary disks and around 25 K in cometary nuclei[11]. Its abundance, therefore, is very sensitive to both the local temperature of the region where comets formed and their thermal history since then. Outside 2.5 au, several comets show coma abundances of CO that exceed that of $H_2O$[12]. $H_2O$ ice starts to sublimate efficiently from comet nuclei within 2.5 au from the Sun, while $CO_2$ and CO ice have lower sublimation temperatures and can effectively sublimate within 13 au and 120 au of the Sun, respectively[13]. The activity of distant comets and Centaurs has therefore been attributed to these hypervolatiles[13]. Inside 2.5 au from the Sun, water sublimation from the nucleus scales approximately with the inverse square of the heliocentric distance and is a good indicator of underlying water abundance. A notable CO-rich comet is C/2016 R2 (PanSTARRS), though with a perihelion of 2.6 au, it never reached the inner solar system where water sublimation is fully effective, and while high $CO^+$ and $N_2^+$ production levels were detected, no evidence of $H_2O$ or its fragments was observed in its coma[14,15], which complicates the comparison with 2I/Borisov. Further complicating the comparison, in C/2016 R2 (PanSTARRS) the dominant volatile is most likely $CO_2$, not CO [16]. For these reasons and uncertainties, we emphasize that the extremely high CO/$H_2O$ abundance measured in 2I/Borisov is significant in comparison to the large number of comets that reach the inner (< 2.5 au) solar system and have measured CO and $H_2O$ values.

Another example of a CO rich comet was Comet C/2009 P1 (Garradd), a highly active, dynamically young comet that was well-observable over an extended time throughout its orbit. When its water production rates decreased after perihelion, its CO production rates increased continuously, leading to a CO abundance of 60% with respect to water at 2 au from the Sun, the highest abundance ratio ever measured[17]. This was interpreted as the result of surface erosion or seasonal illumination effects exposing more layers that reflected the comet's primitive ice composition. Comets C/2009 P1 and 2I/Borisov show similar trends in the CO/$H_2O$ abundance ratio, increasing after perihelion (where mass loss rates and thus surface erosion peak). However, at similar distances from the Sun, 2I/Borisov has a CO/$H_2O$ abundance that is at least three times higher than that of C/2009 P1 (Garradd). Water production rates of 2I/Borisov dropped rapidly after perihelion[9], but our results suggest that CO production rates remained constant or even increased during that period (Figure 2). This could be interpreted as illumination of parts of the nucleus that were obscured during approach, as a chemical heterogeneity of the nucleus, or as the exposure of deeper layers by surface erosion. For comet 67P/Churyumov-Gerasimenko, seasonal and evolutionary effects result in different erosion rates and chemical abundances for its southern and northern hemispheres[18]. As such, the composition of 67P's more eroded southern hemisphere is likely more representative the primitive chemical abundance of the volatiles[19].

Pre-discovery observations of the dust surrounding 2I/Borisov indicated that its activity between 8 and 5 au was likely driven by an ice more volatile than $H_2O$, such as $CO_2$ or CO[20]. However, the observed production rate trends imply that near perihelion $H_2O$ was driving the activity. 2I/Borisov likely has a small nucleus with a radius between 0.2 and 0.5 km[21], and it has been estimated that the comet lost between 1.0 – 6.4 m of its surface, depending on the size of the nucleus[9]. Results from the *Rosetta* mission suggest an orbital thermal skin depth of a few tens of centimetres to 1 meter[22], thus the activity around perihelion for 2I/Borisov would have exposed layers that had not been thermally altered. In this scenario above, the high CO to $H_2O$ abundance of 2I/Borisov is a primordial property that reflects the composition of the ices in the nucleus[23]. This requires that the CO originate from a pristine portion of the object that has remained below the 25 K sublimation point of CO since formation.

Dynamical models of our solar system predicted that large amounts of the comets in our Oort Cloud could be interstellar interlopers[24], but it was unclear what properties would set such objects aside from comets from our solar system. 2I/Borisov's orbit firmly places its origins outside our solar system. Most of its properties, including the colour of its dust, its brightness trend with respect to the distance to the Sun, and the presence of fragment species have been surprisingly similar to those of comets from our solar system[1,9,20,23,25], which could partially be explained by the exposure to radiation during the time the comet spent in interstellar space. Two major differences between typical solar system comets and 2I/Borisov become apparent with our HST/COS observations. First, the high abundance ratio of $CO/H_2O$ measured in 2I/Borisov firmly sets it apart and informs us both of the thermal history and origins of the nucleus, as well as the CO and $H_2O$ ice abundance in the outer reaches of the host system. Second, the elemental abundance of carbon relative to oxygen in 2I/Borisov's volatiles is nearly six times higher than that of previously measured solar system comets (Extended Data Figures 3, 4, and 5), a clear indication that the protoplanetary disk where 2I/Borisov formed was enriched in carbon relative to our own, while exhibiting no other peculiar differences in other elements (Figure 3). Such major differences between previously measured comets and the first interstellar comet invite investigation and explanation.

The previous closest encounter to a star of 2I/Borisov was with Ross 573, an M0V type star, about a million years ago at a distance between 11,000 and 19,000 au[26]. This encounter was not close enough to heat the surface. Beyond this time it is impossible to trace a possible host system of the interstellar comet. However, with the COS discovery of a $CO/H_2O$ ratio greater than 1, and by assuming that 2I is statistically likely to be a representative of typical comets rather than an outlier in its host system, we can place three constraints on 2I/Borisov's origin system. First, the host system must be chemically distinct from our own in order to form a $CO/H_2O$ ratio greater than 1, a ratio not observed in our solar system even among dynamically new comets. Second, 2I/Borisov must have formed beyond a CO snowline of its host system to both form and preserve a $CO/H_2O$ ratio greater than 1 (multiple snowlines are possible, and they move over time [27]). Solar system comets may have formed around the CO snowline, which explains the range of CO abundances observed[11]. Substantial CO ice must have formed concurrently with $H_2O$ ice and cannot be trapped in a clathrate. Third, 2I/Borisov must have formed in a system that experienced dynamical interactions sufficient to strip planetesimals beyond a CO snowline. Such interactions must eject planetesimals with primordial CO and $H_2O$ ice and preserve the CO ice during the encounter[6].

      M-type stars are the most common stars in our galaxy. They form late in their respective stellar neighbourhoods and have low temperatures and luminosities that place the CO snowline at nearly 6.6 au. Such a stellar system would fit the above requirements for 2I/Borisov's host system well.

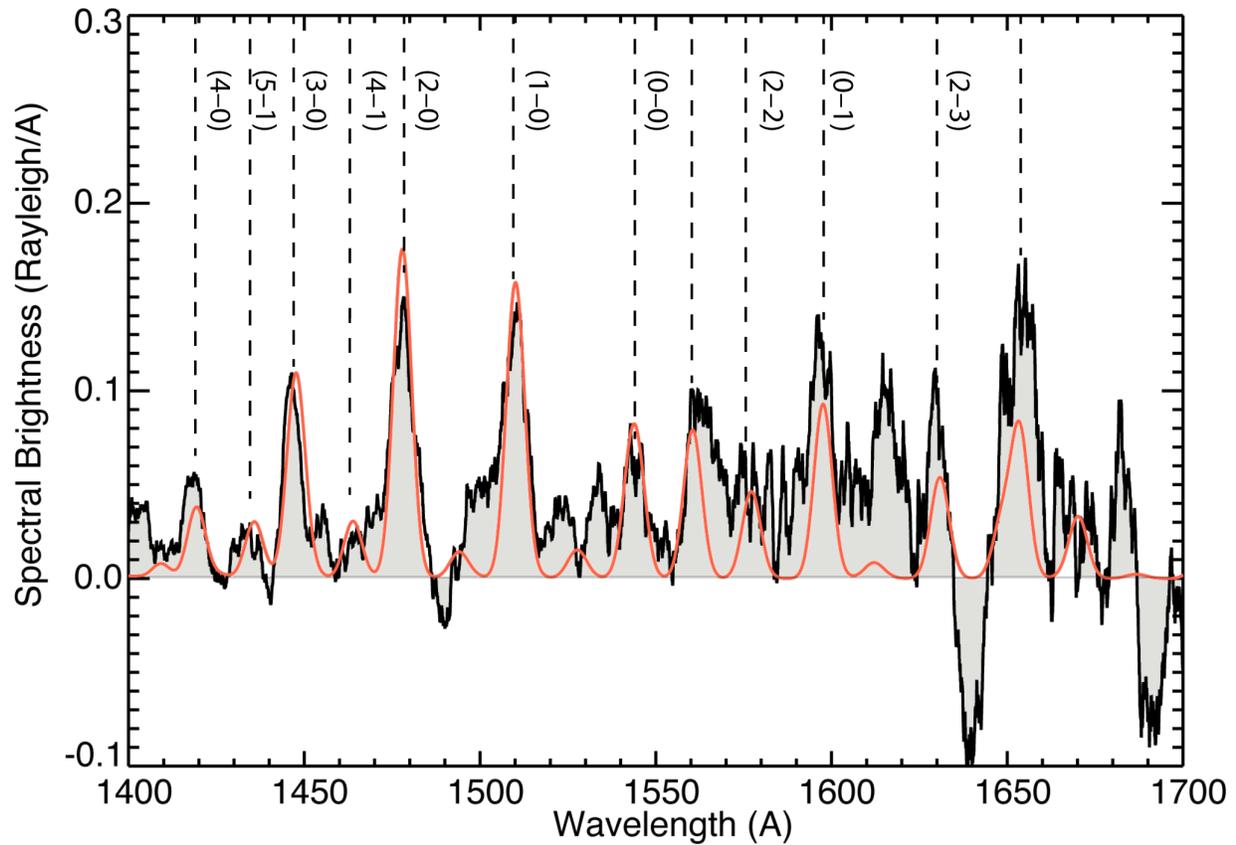

**Figure 1 | Co-added HST/COS spectra of 2I/Borisov acquired between Dec. 19 – 22, 2019.** Black lines indicate the smoothed HST/COS data, and red lines the best-fit CO fluorescence model for the CO Fourth Positive system. The total integration time was 17,901 s. The dashed vertical lines and labels indicate individual transitions of the CO Fourth Positive system[7]. The grey shading shows the area above and below the zero line for the spectral channels.

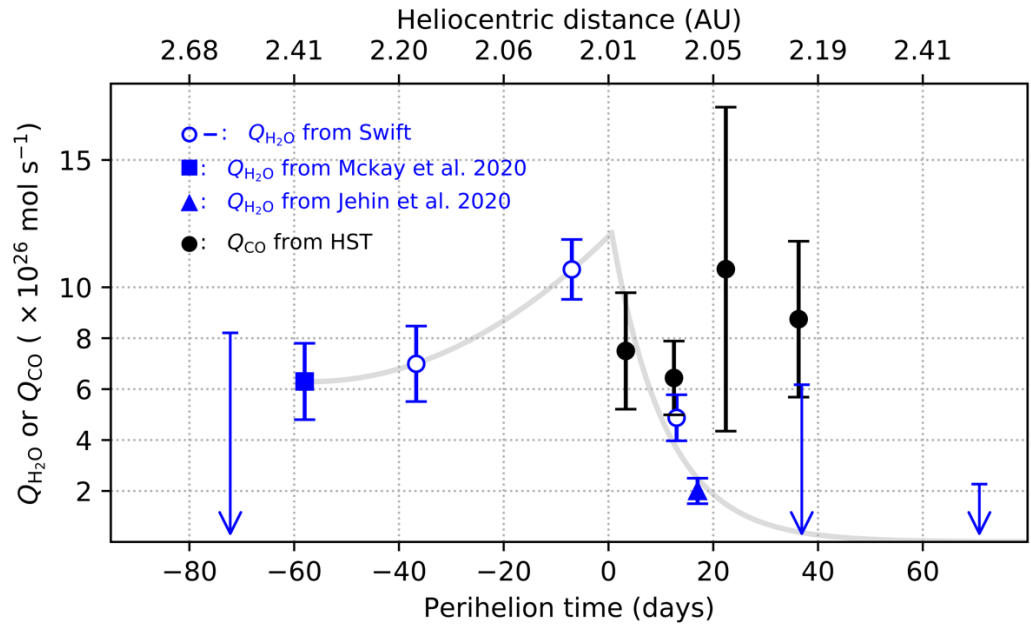

**Figure 2 | Volatile production rates as a function of time relative to perihelion.** The production rates of CO measured with HST/COS (this work) and the water production rate measured by Swift (based on OH, open circles; 9) and the Very Large Telescope/UVES (based on OH, 10), and at the Apache Point Observatory (based on [OI], 8). Arrows indicate 3-σ upper limits, and error bars indicate 1-σ stochastic uncertainties. The grey line indicates the temporal trend of water production rates used to derive the elemental composition.

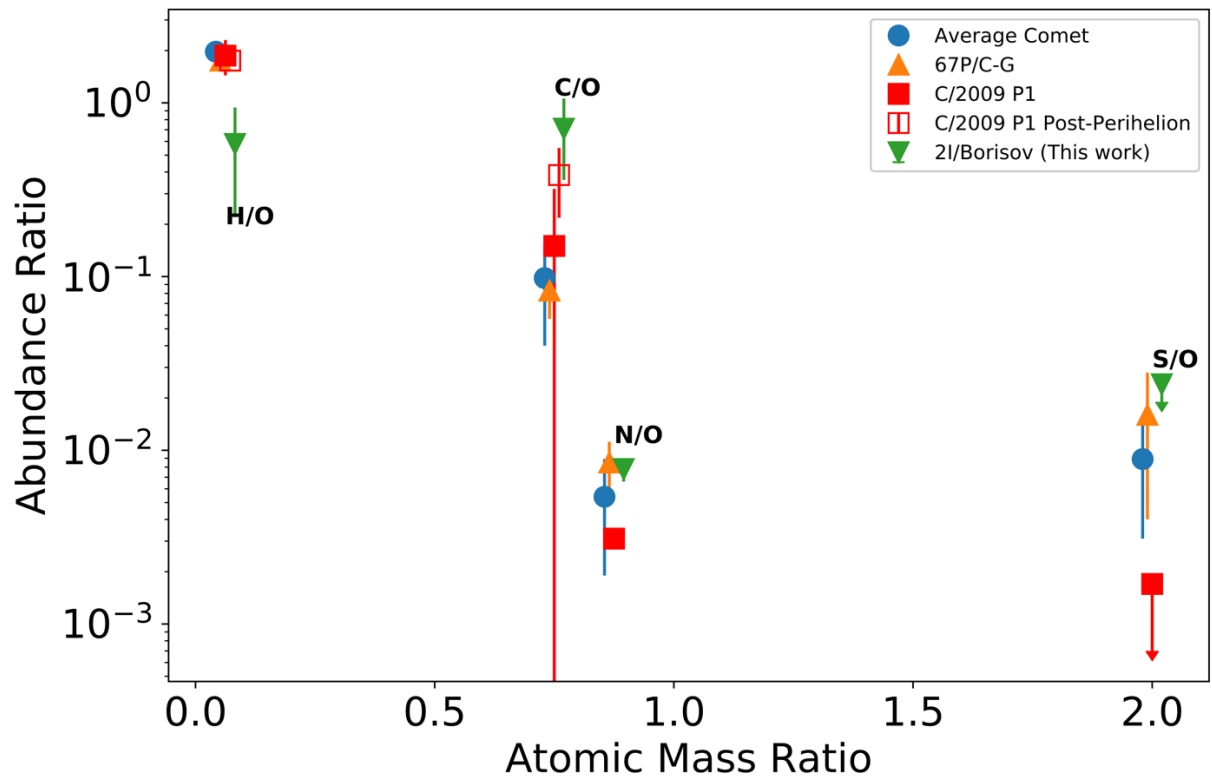

**Figure 3 | The composition of volatiles in the coma of 2I/Borisov compared to comets in our solar system.** The elemental abundance ratio with respect to atomic oxygen are shown for 2I/Borisov, 67P/Churyumov-Gerasimenko [28] and average comets[18]. Averaged elemental ratios[29] and enhanced ratios post-perihelion[17] are shown for comet C/2009 P1 (Garradd). Error bars are 1-σ and upper limits are indicated by a small arrow below the symbol (S/O for Borisov).

**Acknowledgements**

Based on observations with the NASA/ESA Hubble Space Telescope obtained from the Mikulski Archive for Space Telescopes at the Space Telescope Science Institute, which is operated by the Association of Universities for Research in Astronomy, Incorporated, under NASA contract NAS5- 26555. Support for Program number GO-16049 was provided through a grant from the STScI under NASA contract NAS5-26555.

We gratefully acknowledge Alison Vick, David Sahnow, Elizabeth Nance, and Crystal Mannfolk at STScI, for their help with scheduling the challenging HST observations. We further acknowledge the effort of many observers who supplied astrometry of 2I/Borisov to the International Astronomical Union's Minor Planet Center to support the planning of the HST sequences, in particular Emmanuel Jehin, Quan Zhi Ye, Marco Micheli, Dave Tholen, Tim Lister, Karen Meech and Scott Sheppard. Part of this research was conducted at the Jet Propulsion Laboratory, California Institute of Technology, under a contract with NASA.

**Author Contributions:**

**Author information:**

Correspondence and requests for materials should be addressed to dennis@auburn.edu. The authors declare no competing financial interests.

# Methods

**Data reduction**

To characterize the composition of volatiles around 2I/Borisov, we used the G140L/800 mode of HST/COS[30]. This observing mode provides gapless coverage from 800 to 1950 Å with a spectral resolving power between 2500 – 3500.

HST/COS spectral data were processed via the CALCOS pipeline, which produces several calibrated spectral products with units of ergs/cm$^2$/s/Angstrom. For our data analysis the **x1dsum** files, the co-added individual spectra for each visit, were used to maximize signal.

A Python routine utilizing the Astropy package[31] is then used to individually access the spectral data and pointing information contained within the "jitter" files (.jit) for further reduction. The apparition of 2I/Borisov required daytime observations, increasing the airglow contribution to the spectral flux of atomic hydrogen and oxygen emissions. However, there is no terrestrial airglow contribution to the CO Fourth Positive Group emission, so airglow correction is not necessary for our study. However, the atomic hydrogen and oxygen features in the bandpass are heavily contaminated by terrestrial airglow emission and therefore unable to provide diagnostic information on atomic H and O abundances.

Individual **x1dsum** spectra are co-added into 4 stacks, based on temporal proximity. The result is shown in Extended Data Figure 2. We note that data acquired on 30 December 2019 were deemed too temporally separated from any other dataset to be reliably coadded, and as a result, have the lowest signal to noise ratio of the four stacks. Data acquired on December 14, 2019, are omitted due to a guide star acquisition issue and subsequent failed pointing of HST. The resulting spectral fluxes remain in ergs/cm$^2$/s/Angstrom until after co-adding, at which point an erg-to-photon conversion array is generated from the wavelength array of the data. Following this conversion to a spectral photon flux to Rayleighs, a spectral brightness useful for analyzing cometary, airglow, and auroral spectra, is implemented. The conversion requires the solid angle of the COS instrument, which has a diameter of 2.5 arcseconds. The solid angle of the COS is then:

(Equation 1) $\Omega = 4\pi sin(\frac{\theta}{2})^2$

where $\theta$ is the diameter of the COS instrument in radians. To convert the photon spectral flux to a spectral brightness we use the equation:

(Equation 2) $I = \frac{4\pi F_{photon}}{10^6 \Omega}$

, where $F_{photon}$ is in units of photons/cm$^2$/s and the resulting $I$ is in units of Rayleighs. Note that if $F_{photon}$ is in photons/m$^2$/s the factor of $10^6$ must be changed to $10^{10}$. These co-added spectra with units of spectral brightness are then used for the CO modelling.

**CO modelling**

To fit the CO emission, we used the fluorescence and radiative transfer model for the CO Fourth Positive system[7]. This model includes over 15,000 transitions for the CO (A $^1\Pi$ –

X $^1\Sigma$) Fourth Positive system and is combined with a radial outflow model for the distribution of the CO gas around the nucleus. Adjustable parameters include rotational temperature ($T_{rot}$ = 75 K), the outflow velocity of the CO from the nucleus (v = 0.85 $r_h^{-0.5}$ km/s), and solar flux (solar minimum). We smoothed the model data with Gaussian profiles to a spectral resolution of FWHM = 6 Å. The 1.25" radius COS aperture corresponds to a radius of approximately 1800 km at the comet. We modeled the spectrum of the CO Fourth Positive system for an initial average column density of $N_{col} \sim 4 \times 10^{13}$ cm$^{-2}$ for a gas production rate of Q(CO) = 8.25 x 10$^{26}$ molecules/s. We used the 3 strongest features in the spectrum (the 1–0, 2–0, and 3–0 bands; see Figure 1) to scale the model to the observations and found the corresponding column densities and production rates by scaling these accordingly.

Fits were completed without referencing the statistical errors computed from the CALCOS pipeline, which has previously been documented having difficulty correctly calculating low signal-to-noise ratio errors[32].

**Uncertainties in the CO production rates**
To estimate statistical uncertainties, we binned the observed spectra by a factor of 30 and put them and the models on the same wavelength grid. We evaluated the variance in the differences between the model and data for the strongest three features for every observation, and multiplied the variances by the FWHM of 6.0 Å to 1-σ estimate absolute errors in the flux, which are respectively 0.37 R, 0.22 R, 0.67 R, and 0.35 R for the four observations. In order to estimate relative errors and derive uncertainties in production rates, for every observation we calculated the observed integrated brightnesses of the three strongest features and derive their mean value, and the ratio between the corresponding absolute error and this mean brightness is regarded as 1-σ relative uncertainty, whose results are 30.5%, 22.5%, 59.4% and 35.0% respectively for the four observations.

There are additional systematic uncertainties both from the model such as the assumption of spherical symmetry and from the data such as neglect of vignetting near the edge of COS aperture caused by the aberrated prime focus of HST and minimal optical elements in COS. This vignetting decreases the throughput of COS with increasing distance from the aperture center.

**Elemental composition of coma volatiles**
The elemental composition of the gases in the coma of each of the comets illustrated in Figure 3 is determined based on the reported abundance of species relative to water:

$$\frac{C}{O} = \frac{X_{CO}}{X_{H2O} + X_{CO}}$$
$$\frac{H}{O} = \frac{2X_{H2O}}{X_{CO} + 2X_{H2O}}$$

For elements that have multiple parents, this ratio is calculated by summing the abundance, *X*, of each molecule, *i*, that contains the element of interest multiplied by the number of atoms of that element in the molecule. For example, S/O would be calculated in a comet with reported abundances of CO$_2$, OCS, and S$_2$ relative to water as follows

$$\frac{S}{O} = \frac{X_{OCS} + 2X_{S2}}{X_{H2O} + 2X_{CO2} + X_{OCS}}$$

The comet abundances used to determine the average of 14 comets in our solar system (Avg. comets in Fig. 3) are based on the compilation of published abundances of 26 different molecules[18]. The uncertainty in these values is based on the range of elemental abundances for those comets with sufficient information to provide an abundance of an element. The elemental abundances of 67P/Churyumov-Gerasimenko are based on the elemental abundances derived using the outgassing rate of numerous species observed by the *Rosetta* Orbiter Spectrometer for Ion and Neutral Analysis (ROSINA) Double Focusing Mass Spectrometer shortly before perihelion[28]. Uncertainties here are based on the uncertainties provided in [28]. Two separate values are reported for C/2009 P1 (Garradd), the elemental abundances H/O and C/O for the time period when the CO abundance was highest and the overall average prior to the increased outgassing of CO[17, 29].

Finally, the elemental abundances and uncertainties in the elemental abundances of 2I/Borisov were determined using a combination of published outgassing rates for CO as reported here, reports of CN, $NH_2$, and $C_2$[23], relative to multiple reports of the water outgassing rates[8,9,10] (which were extrapolated to the time of each relevant observation). The production rates are summarized in Extended Data Figure 4. We extrapolated the pre-perihelion water outgassing rate by fitting a two-degree polynomial to the three data points prior to perihelion on 11 October, 1 November, and 1 December 2019:

Pre-perihelion: $Q_{H2O}$ = 1.77e23 * $\Delta T^2$ + 2.01e25 * $\Delta T$ + 1.2e27 (molecules/s)
Post-perihelion: $Q_{H2O}$ = exp(62.42184 – 0.0964962*$\Delta T$) (molecules/s)

, where DT is the number of days from perihelion (negative before and positive after).

We then determined the peak outgassing rate at perihelion based on this equation and fit an exponential equation to the estimated peak production rate and the two measured post-perihelion production rates from 21 and 25 December 2019. All measurements of CN, $NH_2$, and $C_2$ took place pre-perihelion and prior to the detection of CO reported here and contribute little to the bulk elemental composition. Therefore, H/O and C/O are determined using the CO observations reported here and the water outgassing rate determined by our equation for extrapolating the outgassing rate on 12, 19, and 30 December 2019, and on 13 January 2020. For both elemental abundances in Fig. 3, we use the average of the four elemental abundances determined using the outgassing rates on those dates (Extended Data Figure 4). The uncertainty is based on the range of measurements. The N/O is determined based on CN and $NH_2$ outgassing rate relative to water using the average for the two dates when both CN and $NH_2$ abundances are available and the uncertainty based on the range of possible values. However, this could be considered an upper limit because CO, a major carrier of oxygen, is not available for this calculation. The determination of the upper limit for S/O is described below. The elemental abundances are summarized in Extended Data Figure 5.

**Sulphur upper limits**

The upper limit and associated error for the production rate of sulfur was determined by subtracting the best-fit CO column density model for the 12/19 – 22 co-added spectrum and searching for SI 1425 Å doublet emission, SI 1475 Å triplet emission, and SI 1813 Å triplet emission. The associated fluorescence efficiencies, or g-factors, for each emission feature were calculated using the atomic properties available on the NIST spectral database (available at https://physics.nist.gov/PhysRefData/ASD/lines_form.html), and a SORCE SOLSTICE high resolution UV solar spectrum from the LISIRD database (available at https://lasp.colorado.edu/lisird/data/sorce_solstice_ssi_high_res/) for the relevant date. We find that the total g-factors for the SI 1425, 1475, and 1813 Å features at 100 K, a heliocentric distance of 2.03 au, and heliocentric radial velocity of 5.05 km/s are 2.83 x $10^{-8}$, 2.62 x $10^{-8}$, and 2.01 x $10^{-6}$ photons/molecule/s. Due to this large difference in g-factors we selected the SI 1813 Å triplet to calculate an upper limit, despite the increased noise in the redder portion of the spectrum. We note that for the 1813 Å triplet the 1807 Å feature is nearly three orders of magnitude larger than the following 1820 Å feature with the input parameters. However, at the resolution of the G140L grating these features are blended.

We then created a synthetic SI spectrum for a range of sulphur column densities, which we then compared to the CO-model subtracted co-added spectrum. We note that our model assumes an optically thin coma, which we adopt in the absence of any emission line profiles to inform optical thickness or radiative transfer models. To estimate uncertainties in the low signal-to-noise NUV portion of the spectrum we also compared synthetic sulphur spectra that would fit within the 1-σ standard deviation error of emissions at wavelengths greater than 1700 Å. The resulting spectra are shown in Extended Data Figure 3. We find an upper limit column density of n(S) < 2±1x$10^{12}$ cm$^{-2}$. Uncertainties in column density dominate uncertainties in g-factors after this stage and are propagated through.

Two Haser models were implemented to determine Q(S); one a simple single component fountain model to determine Q(S) directly, the other a two-component model assuming $CS_2$ as the parent molecule of sulphur. Lifetimes for sulphur and $CS_2$ at 1 au[33] and were scaled to a heliocentric distance of 2.03 au. The velocity of $CS_2$ is assumed to be 0.85 $r_H^{-0.5}$ km/s, and the velocity of S is assumed to be 1.3 $r_H^{-0.5}$ km/s, where $r_H$ is heliocentric distance. By dividing the upper limit for column density by the integrated Haser model for number of molecules in the COS field of view (1800 km in radius at 2I/Borisov), an upper limit for Q(S) and Q($CS_2$) can be found. We find that Q(S) < 7x$10^{24}$ mol/s and Q($CS_2$)< 1.4x$10^{25}$ molecules/s. We can then calculate the S/O with either of the two upper limits. For the case with a larger upper limit, where $CS_2$ is the sole source of S, there are two sulphur atoms per molecule for $CS_2$ and one oxygen atom apiece for CO and $H_2O$, S/O = Q($CS_2$)*2/(Q(CO)+Q($H_2O$)), and therefore S/O < 0.026 ± 0.013. For the general sulphur case, Q(S), S/O = Q(S)/(Q(CO)+Q($H_2O$)), we find S/O < 0.009±0.005. We therefore place a conservative 3-σ upper limit on S/O of 0.024.

**Data availability**

All data are publicly available on the Mikulski Archive for Space Telescopes under HST proposal 16049, PI Dennis Bodewits.

# Extended Data

**Extended Data Figure 1 |** Observing logs of HST observations and water production rates from simultaneous observations with Swift/UVOT (Xing et al. 2020). $T-T_p$ indicates the time past perihelion; $R_h$ indicates the heliocentric distance and $dR_h$ the comet's radial heliocentric velocity; $\Delta$ indicates the distance to Earth; $N_{col}$ is the derived average column density in the field of view; $Q(CO)$ and $Q(H_2O)$ indicate the CO and water production rates corresponding to these column densities; and the *radius* is the radius of the field of view at the distance of the comet.

| Date UTC | $T-T_p$ (days) | $R_h$ (au) | $dR_h$ (km/s) | $\Delta$ (au) | $T_{exp}$ (s) | $N_{col}$ ($10^{13}$ cm$^{-2}$) | Q(CO) ($10^{26}$ molec/s) | Q(H$_2$O) ($10^{26}$ molec/s) | Radius (km) |
|---|---|---|---|---|---|---|---|---|---|
| 12/11/2019 | 3.3 | 2.01 | 2.04 | 1.98 | 4105 | 3.5 ± 1.1 | 7.5 ± 2.3 | | 1795 |
| 12/19-22/2019 | 12.5 | 2.03 | 5.50 | 1.94 | 17901 | 3.0 ± 0.7 | 6.4 ± 1.4 | 4.9 ± 0.9 | 1759 |
| 12/30/2019 | 22.4 | 2.07 | 9.26 | 1.94 | 1704 | 5.0 ± 3.9 | 10.7 ± 6.4 | | 1759 |
| 1/13/2020 | 36.3 | 2.16 | 14.25 | 1.97 | 5166 | 4.1 ± 1.5 | 8.7 ± 3.1 | <5.6 | 1786 |

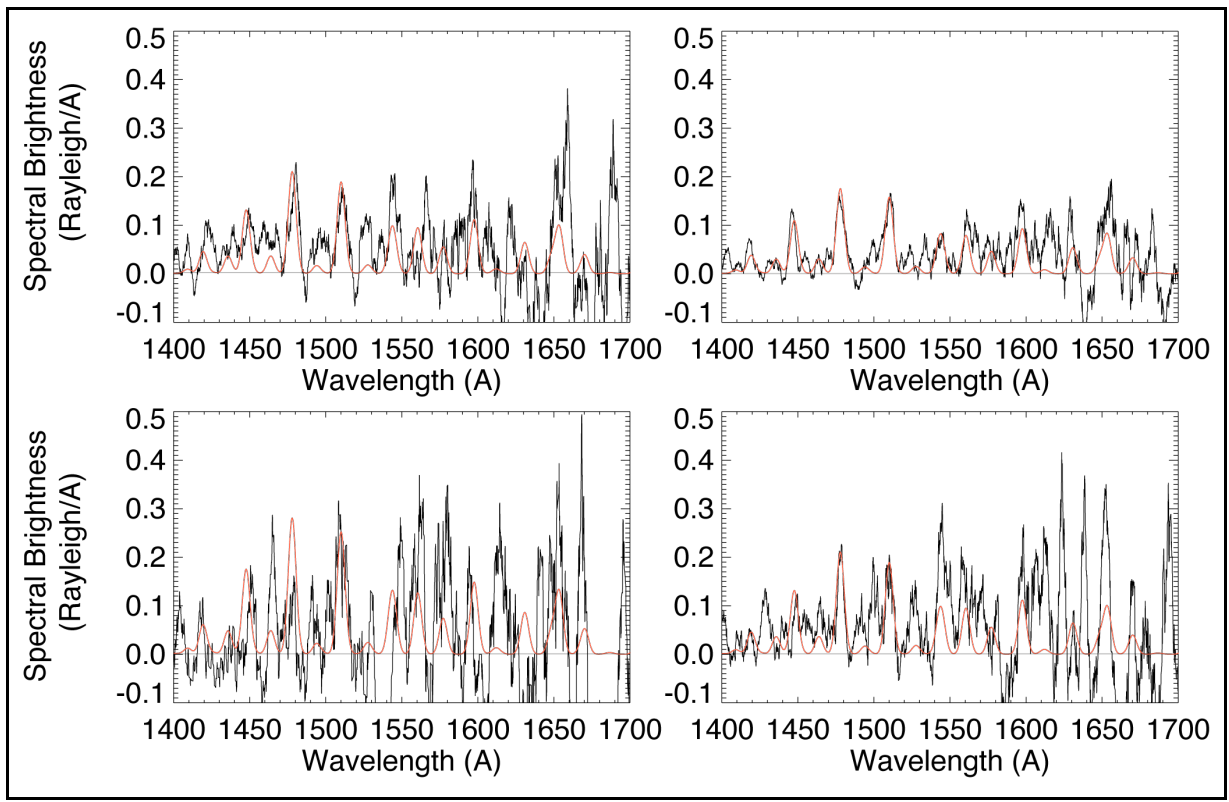

**Extended Data Figure 2 | Co-added HST/COS spectra of 2I/Borisov.** Top left, Dec. 11-14, 2019 UTC; Top right, Dec. 19-22; Bottom left, Dec. 30, 2019; Bottom right, Jan. 13, 2020. Black lines indicate the smoothed HST/COS data, and red lines the best-fit CO fluorescence model.

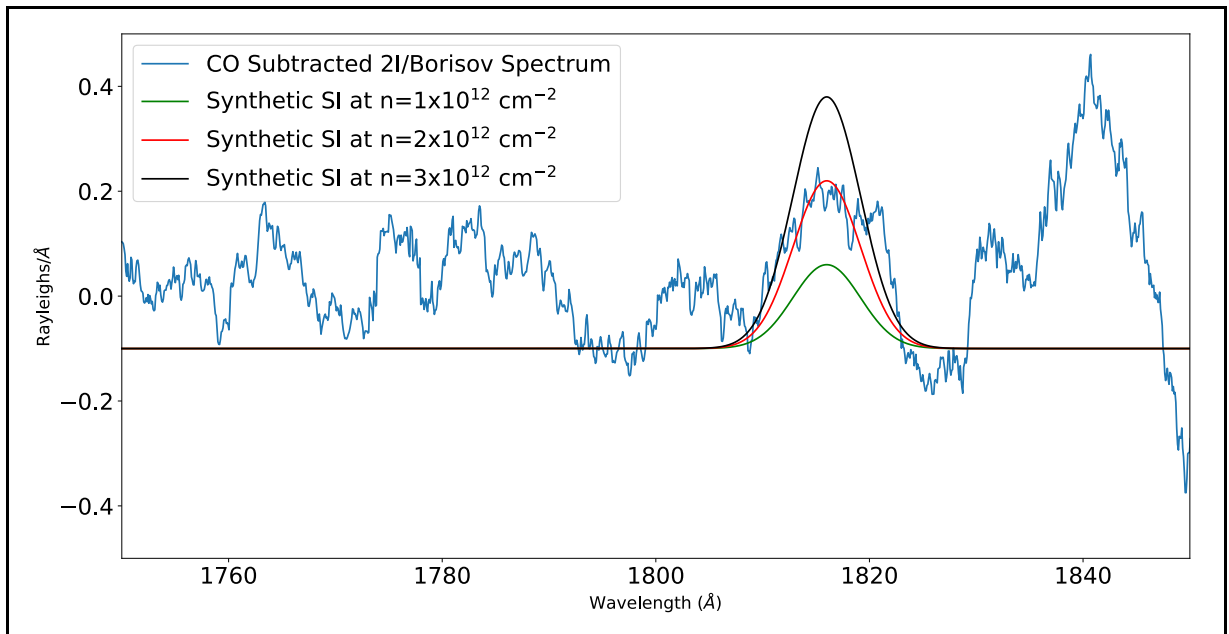

**Extended Data Figure 3 | Upper Limits on Sulphur in 2I/Borisov.** Synthetic atomic sulphur blended triplet emissions at 1813 Å plotted for three different column densities of sulphur. The filled region of the spectrum represents one standard deviation.

**Extended Data Figure 4 | Production rates used to determine the elemental composition of volatiles in the comae of 2I/Borisov.**

| Species | Date | Production Rate | Reference |
|---|---|---|---|
| | (UTC) | (Molecules/s) | |
| $NH_2$ | Nov. 5, 2019<br>Nov. 25, 2019 | 4.2e24<br>4.8e24 | 23<br>23 |
| CN | Nov. 5, 2019<br>Nov. 25, 2019 | 2.4e24<br>1.6e24 | 23<br>23 |
| CO | (see ED Table 1) | | This work |
| $H_2O$ | Nov. 5, 2019<br>Nov. 25, 2019<br>(see ED Table 1 for dates for comparison to CO) | 7.29e26<br>9.69e26 | 8,9,10 |
| $CS_2$ | Dec. 19-22, 2020 | $< 1.4 \times 10^{25}$ | This work |

**Extended Data Figure 5 | Elemental composition of volatiles in the comae of different comets compared with 2I/Borisov.** See text for references.

| Object | H/O | C/O | N/O | S/O |
|---|---|---|---|---|
| Avg. comets | 1.97±0.13 | 0.098±0.058 | 0.0054±0.0035 | 0.0089±0.0058 |
| 67P/C-G coma | 1.75±0.11 | 0.083±0.026 | 0.0085±0.0027 | 0.016±0.012 |
| Garradd Avg. | 1.87±0.43 | 0.15±0.03 | 0.0031±0.0003 | < 0.0017 |
| Garradd Post-perihelion | 1.14±0.49 | 0.384±0.166 | | |
| Borisov | 0.58±0.36 | 0.71±0.35 | 0.0078±0.0012 | <0.009±0.005 |